\documentclass[9pt,twocolumn,twoside]{osajnl}
\journal{optica}
\usepackage{amssymb}    
\usepackage{mathptmx}  
\usepackage{amsmath} 
\usepackage{mathrsfs}
\usepackage{fancyhdr} 
\setboolean{shortarticle}{false}
\title{Single-Axis Ptychographic Coherent Diffractive Imaging for Spectroscopic and Wavefront Retrieval}

\author[1]{Qijun You}
\author[1]{Lingshuo Meng}
\author[1]{Fangrui Quan}
\author[1,*]{Wei Cao}

\affil[1]{School of Physics and Wuhan National Laboratory for Optoelectronics, Huazhong University of Science and Technology, Wuhan 430074, China}
\affil[*]{Corresponding author: weicao@hust.edu.cn}

\begin{abstract}
	
	We present a novel axial ptychographic coherent diffractive imaging (AP-CDI) technique designed to overcome the critical throughput bottleneck of conventional methods. By replacing the 2D raster scan with a simple 1D axial scan, our approach reduces the number of required diffraction patterns by approximately an order of magnitude while maintaining high-fidelity reconstruction. We have experimentally validated this concept, successfully performing simultaneous spectroscopic imaging of a sample and quantitative wavefront characterization of the illumination, thereby accurately quantifying the chromatic aberration of a broadband field. This capability establishes AP-CDI as a highly efficient and versatile tool for real-time, multi-modal imaging, with immediate potential in ultrafast science, material characterization, and live-cell bio-imaging.
\end{abstract}
\setboolean{displaycopyright}{true}

\begin{document}
	\maketitle
	\section{Introduction}
  Lensless imaging has emerged as a transformative paradigm in microscopy, circumventing the physical constraints of conventional optics by digitally reconstructing a sample’s diffracted wavefront \cite{Aidukas2024, Daurer:21}. In essence, the technique replaces imaging lenses with iterative phase retrieval algorithms \cite{ph2,ph3,ph7,ph8}, enabling recovery of complex amplitude distributions from far-field diffraction patterns under well-defined measurement constraints. This computationally driven architecture offers a compact, alignment-free, and cost-efficient platform for quantitative wavefront reconstruction. Owing to these advantages, lensless imaging has shown substantial promise in biomedical diagnostics, particularly in settings where simplicity and portability are critical. Biological specimens such as cells and thin tissue sections generally exhibit negligible absorption in the visible range but induce phase modulations that encode rich morphological and biochemical information. Quantitative phase imaging (QPI) leverages this property to retrieve parameters such as cell thickness, refractive index, and dry mass without exogenous labeling, thereby enabling label-free, noninvasive, and long-term observation of cellular dynamics and structure–function relationships \cite{bio1, bio2}. Despite these advances, conventional lensless imaging typically depends on highly coherent and monochromatic illumination, which fundamentally limits spectral information retrieval and impedes its applicability to multispectral biological characterization. Multispectral imaging, by contrast, reflects the biochemical and functional diversity of biological tissues, providing essential information for comprehensive diagnosis.
  
  To overcome these limitations, Ptychographic Interferometric Multiplexing (PIM) \cite{broadband3} has been developed as a powerful computational strategy capable of reconstructing multiple spectral components simultaneously within a partially coherent source. This approach effectively bridges the gap between lensless and multispectral imaging. Conceptually, PIM extends the framework of ptychography—a well-established lensless imaging technique that reconstructs objects from highly redundant diffraction measurements acquired via spatially overlapped probe illumination \cite{ptychography1, ptychography2}. Over the past decade, ptychography-based multispectral lensless imaging schemes have demonstrated exceptional performance across both visible and X-ray regimes \cite{broadband2,broadband5,broadband7,Du23, broadband8,broadband10}. However, these techniques generally rely on two-dimensional scanning architectures that substantially increase system complexity, require costly precision stages, and impose stringent stability requirements. Moreover, the acquisition of large volumes of redundant diffraction data leads to heavy computational burdens and prolonged reconstruction times, hindering deployment in real-time, high-throughput, or field-portable imaging applications.
  
  	\begin{figure}[ht]
  	\centering\includegraphics[width=1\linewidth]{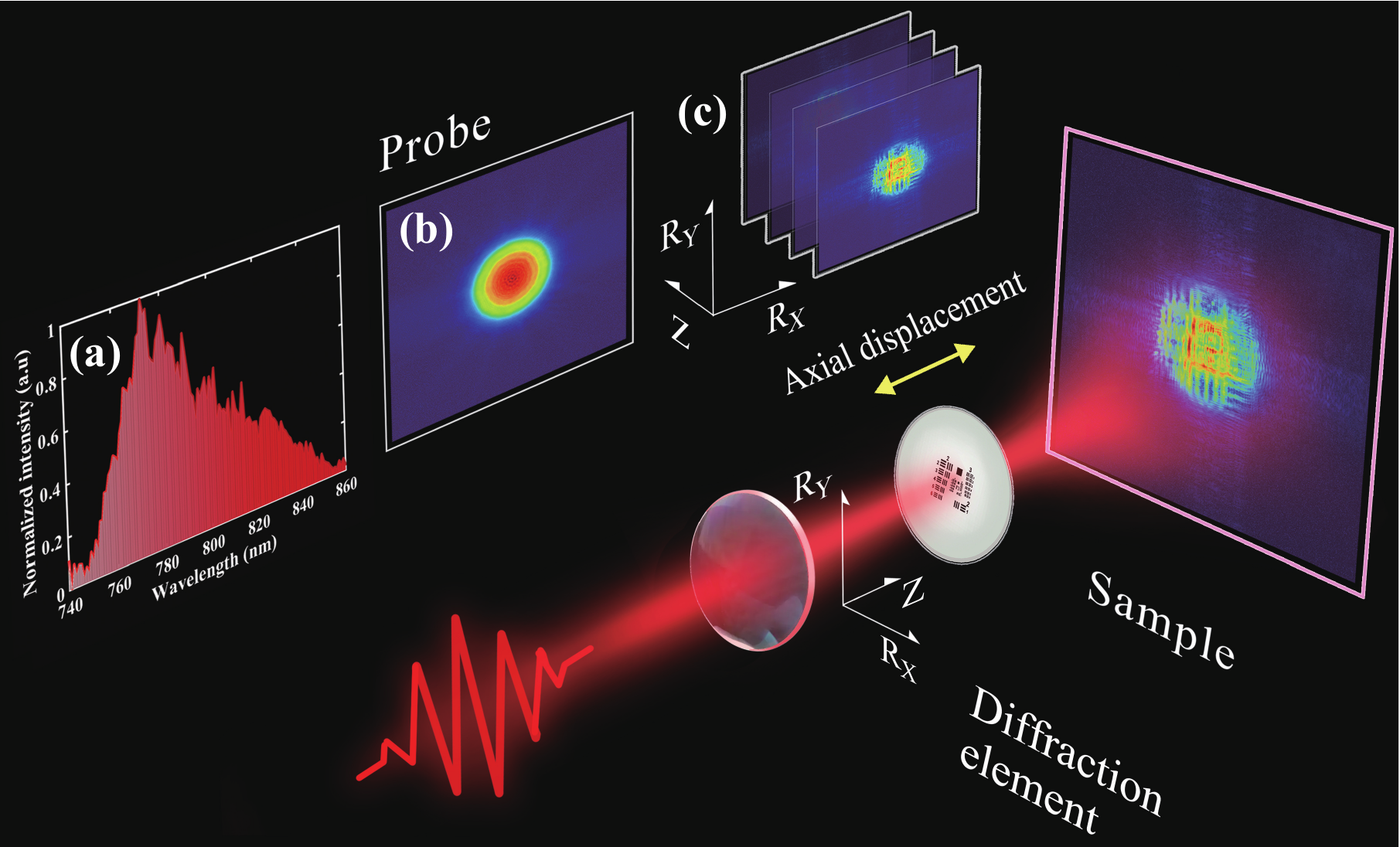}
  	\caption{\label{fig1} \footnotesize Axial ptychographic imaging system. The sample is mounted on a motorized translation stage and moved incrementally along the optical axis. The beam sequentially illuminates the sample at each position to generate exit waves, while the detector simultaneously acquires the corresponding diffraction patterns. (a) Spectrum of the femtosecond source. (b) Diffraction data of the probe beam acquired without the sample. (c) Diffraction data set acquired during axial movement of the sample.}
    \end{figure}

   \begin{figure*}[ht]
	\centering\includegraphics[width=1\linewidth]{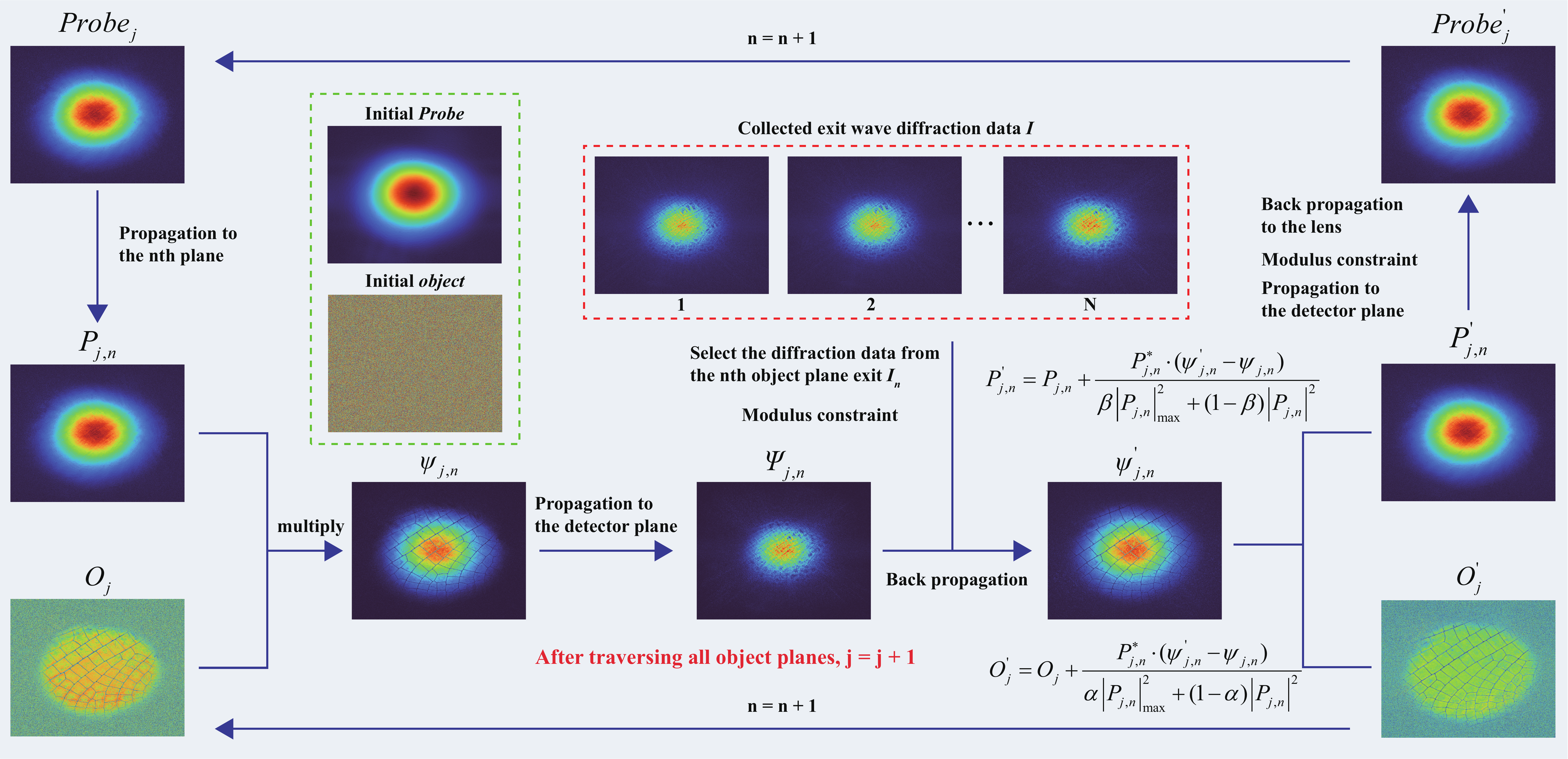}
	\caption{\label{fig2}\footnotesize Axial Phase Retrieval Algorithm. Each iteration of the algorithm comprises $N$ cycles executing this process, successively updating the estimates of the object and probe by utilising diffraction patterns from exits on different object planes.}						
   \end{figure*}

   In response, a novel multispectral lensless imaging approach based on axial scanning is introduced here. Unlike conventional two-dimensional schemes, this technique employs single-axis translation of the sample to acquire partially coherent diffraction patterns. By leveraging the inherent redundancy in axial diffraction data, it enables efficient reconstruction of both the sample function and the probe beam. The design significantly reduces system complexity, decreases requirements for data volume and optical stability, and allows simultaneous reconstruction of multi-wavelength information \cite{Md, Md2, You}. Furthermore, the method exhibits excellent wavefront reconstruction capabilities, enabling precise characterization of the phase distribution of various optical elements—including conventional lenses, diffractive optical elements, and metalenses \cite{metalense1, metalense2, metalense3,metalense4}. Such functionality offers new avenues for studying wavefront aberrations, such as chromatic aberration correction and dispersion analysis, with potential applications in adaptive optical systems and precision optical metrology. Compatibility with partially coherent sources also suggests broad potential in areas such as broadband optical system characterization and femtosecond laser pulse shaping.
  
  Based on single-axis scanning and an information-multiplexed phase retrieval algorithm, this work presents a versatile platform compatible with partially coherent sources, combining label-free biological multispectral imaging with optical wavefront diagnostics. This technology not only achieves high-resolution phase characterization but also significantly reduces implementation costs through system simplification, opening new pathways for the practical application of computational microscopy. This breakthrough is expected to have a substantial impact across various fields, including biomedical research, optical engineering, and precision measurement.
  
     \maketitle
     \section{Result}
     
   The core challenge in phase measurement technology lies in recovering missing phase information from captured intensity images. Common approaches involve encoding the input light field to transform phase information into recordable intensity variations, subsequently restoring phase data through decoding \cite{encode1, encode2, encode3}. However, for illumination-encoded systems, differing encoded illuminations may induce scattering effects in the sample, leading to variations in measured optical responses under distinct illumination conditions. Furthermore, employing measurement-encoding techniques—such as designing encoded masks in front of the camera—entails light intensity loss and increases imaging system costs.
   
   Lensless imaging approaches based on multi-distance phase retrieval (MDPR) design calculate phase information by measuring intensity distributions at different positions on the sample's exit plane \cite{metalense1, YOU2}. However, this inherently reconstructs the sample's exit light field under illumination, failing to separate the illumination function from the sample function. More critically, existing methods cannot quantitatively characterise the phase of partially coherent light fields containing multiple frequency components.
   
   To address existing challenges, we have designed a multispectral imaging system with quantitative phase characterisation capabilities and proposed a corresponding phase retrieval algorithm. As illustrated in Fig.~\ref{fig1}, the sample is positioned on a motorized translation stage for axial scanning, with the detector capturing the intensity of the scattered light field generated when the sample is illuminated. At different positions, the sample is illuminated by light fields with varying wavefronts, thereby providing sufficient redundant information. This enables not only the reconstruction of the illumination function and the sample's transmission function, but also the simultaneous recovery of multiple frequency components within the partially coherent field.
   
   \subsection{phase retrieval algorithm}
 The relationship between the directly measured light intensity distribution $I$ and the target light field distribution $\psi$ can be expressed in vectorized form through the forward model \cite{CCTV1,CCTV2}:
 
 \begin{equation}\label{eq1}
 	I=\left| A \psi \right| ^2
 \end{equation}
 
  The transfer matrix $A$, determined by the measurement system, is modeled using the angular spectrum propagation method in our algorithm. It can be further decomposed as $A = F^{-1} H F$, where $F$ and $F^{-1}$ denote the Fourier transform and inverse Fourier transform, respectively. The transfer function at propagation distance $ z $ is given by~\cite{Fourier}:
 
 \begin{equation}\label{eq2}
 	H_{z}(f_x, f_y) =
 	\begin{cases}
 		\exp\left(i2\pi z \sqrt{1 - \lambda^2 (f_x^2 + f_y^2)}\right), & f_x^2 + f_y^2 \leq \frac{1}{\lambda^2} \\
 		0, & \text{otherwise}
 	\end{cases}
 \end{equation}
 
 where $f_x$ and $f_y$ denote spatial frequency coordinates. To ensure consistency with the forward model in Eq.~\ref{eq1}, we define the fidelity function as $F(\psi) = \frac{1}{2}  \| |A \psi | - \sqrt{I} \|_2^2 $. Based on this forward model \cite{error1,error2}, The phase retrieval problem is formulated as minimizing the fidelity function $ \min_{\psi} \frac{1}{2} \| |A \psi| - \sqrt{I} \|_2^2 $.
 
 Using gradient descent with Wirtinger calculus for the complex-valued variable $ \psi $, the gradient is derived as $\psi^{'} = \psi - \gamma \nabla_\psi F(\psi)$.
 Setting the step size $\gamma = 2$ yields the simplified update rule: 
 
  \begin{equation}\label{eq3}
 	\psi^{'} = A^H \operatorname{diag} \left( \frac{A \psi}{|A \psi|} \right) \sqrt{I}
 \end{equation}

%
%
%
 This formula is equivalent to propagating the exit wave to the detector plane using the angular spectrum method while preserving its phase distribution. The calculated intensity is then replaced with the measured value $\sqrt{I}$, and the updated far-field light field is propagated back to the near-field plane. This process is termed  modulus constraint. In this work, the light field $\psi_n(\mathbf{r}) = P_n(\mathbf{r}) \cdot O(\mathbf{r})$ denotes the exit wave at the $n$th plane, where $O(\mathbf{r})$ is the transmission function of the sample, $P_n(\mathbf{r})$ denotes the incident wave at the $n$th plane, where $\mathbf{r} = (x, y)$ represents the two-dimensional coordinate system of that plane. In the near field, a fidelity function may likewise be defined:
 
  \begin{equation}\label{eq4}
 	R(O)=\| \psi^{'}_n (\mathbf{r}) - P_n(\mathbf{r}) \cdot O(\mathbf{r})  \|_2^2	
  \end{equation}
   
  We aim to identify modifications to the object function that reduce this error, thereby bringing the output waveform $P_n(r) O(r)$ closer to the updated output waveform derived from Eq.~\ref{eq3}. According to the error gradient descent method, the update of the object function can be written as:
    
   \begin{equation}\label{eq5}
  	O^{'}(\mathbf{r}) = O(\mathbf{r}) -\frac{\zeta}{2} \nabla R(O) = O(\mathbf{r}) + \zeta P_n^*(\mathbf{r}) \cdot 	[  \psi^{'}_n(\mathbf{r})-\psi_n(\mathbf{r}) ]
   \end{equation}
  
    Let $\zeta$ denote the indeterminate stride satisfying $\min R(O^{'})$, then it follows readily that $\zeta=1/\left| P_n(\mathbf{r}) \right|^2 $.
  
    A weighting function $\alpha \left| P_n(\mathbf{r}) \right|^2 / \left| P_n(\mathbf{r}) \right|^2_{\operatorname{max}}$ was employed, varying spatially in intensity, approaching 1 in the detector's bright regions and 0 in its dark regions \cite{epie}. The variable stride can be denoted as $ \alpha/\left| P_n(\mathbf{r}) \right|_ {\operatorname{ max }}^2 $. $\alpha$ is a constant with a value between 0 and 1. Combining with Eq.~\ref{eq5}, we can obtain the update for the object function:
  
   \begin{equation}\label{eq6}
  	O^{'}(\mathbf{r}) = O(\mathbf{r}) +  \alpha \frac{ P_n^*(\mathbf{r}) }{ \left| P_n(\mathbf{r}) \right|_ {\operatorname{ max}}^2 } \cdot [  \psi^{'}_n(\mathbf{r})-\psi_n(\mathbf{r}) ]
   \end{equation}

    Similarly, we can derive the update for the probe function:
   
    \begin{equation}\label{eq7}
  	P_n^{'}(\mathbf{r}) = P_n(\mathbf{r}) +  \beta \frac{ O^* (\mathbf{r}) }{ \left| O(\mathbf{r}) \right|_ {\operatorname{ max}}^2 } \cdot [  \psi^{'}_n(\mathbf{r})-\psi_n(\mathbf{r}) ]
    \end{equation}

    $\beta$ is a constant with a value between 0 and 1. By introducing regularization constraints, more robust and efficient reconstruction can be achieved \cite{rpie}. 
%

   \subsection{quasi-monochromatic experiment results}
    Taking a quasi-monochromatic source as an example, the specific workflow of the proposed phase retrieval algorithm is illustrated in Fig.~\ref{fig2}. It should be noted that the scattering image of the probe must be captured once before acquiring the scattering images of the sample. Provided that the illumination source remains unchanged, subsequent experiments involving different samples do not require re-acquisition of the probe image. The acquired probe scattering image is used in the algorithm to impose an  modulus constraint on the illumination field (see Methods section for details).
    
   \begin{figure}[ht]
  	\centering\includegraphics[width=1\linewidth]{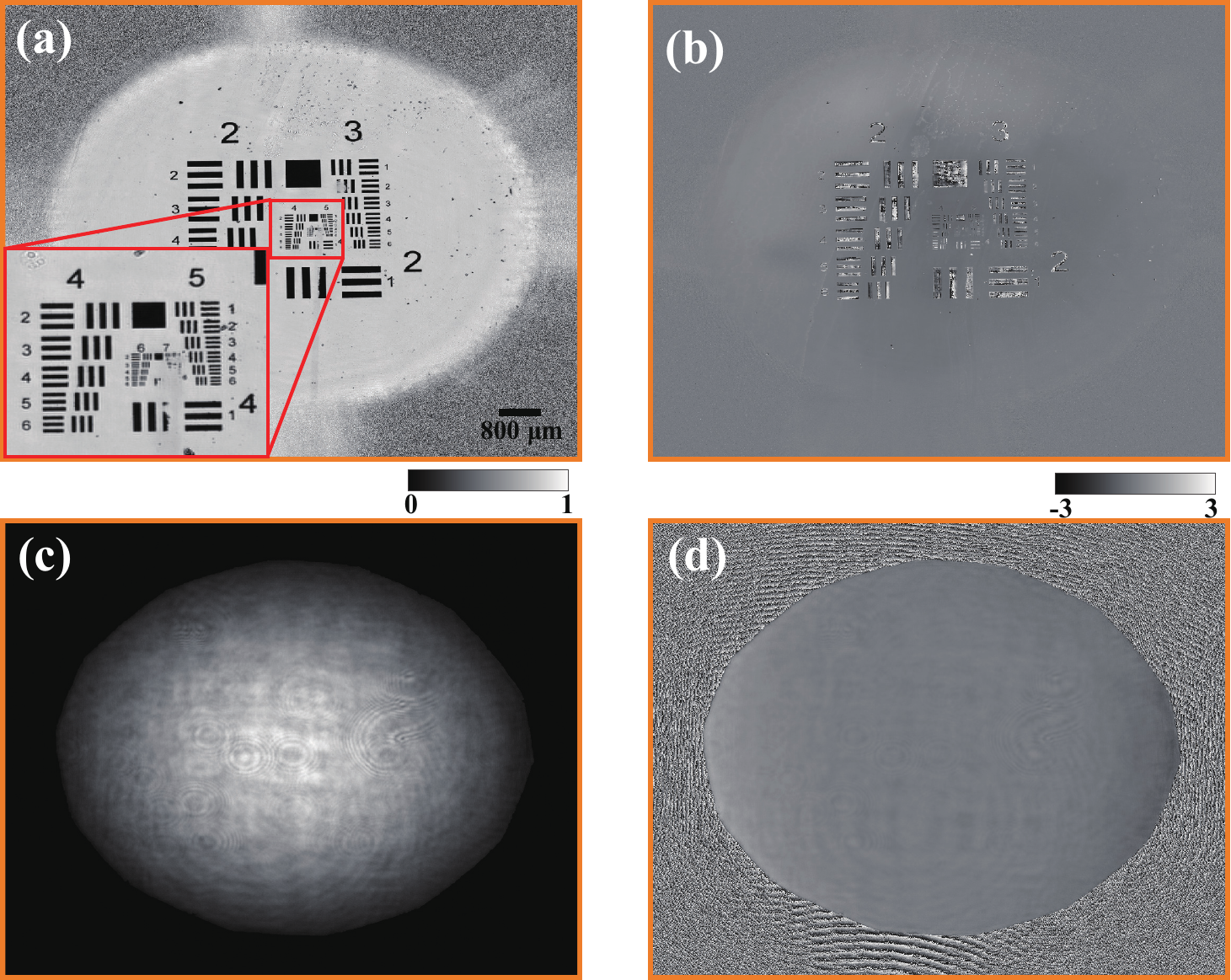}
  	\caption{\label{fig3} \footnotesize Experimental results obtained using the USAF 1951 standard resolution test chart as the sample under quasi-monochromatic source conditions. (a) Amplitude of the reconstructed object transmission function, with the enlarged red region showing more details of the sample. (b) Phase of the reconstructed object transmission function. (c) Amplitude of the reconstructed probe function at the aperture plane, where contamination of the optical elements causes the reconstructed spot to deviate from an ideal Gaussian profile. (d) Phase of the reconstructed probe function at the aperture plane, which deviates from an ideal plane wave due to contamination. }
   \end{figure}  

    We conducted experimental validation using the USAF 1951 standard resolution test pattern as the sample. A semiconductor laser with a wavelength of 532 nm served as the light source, whose beam sequentially passed through a precision pinhole (50 $\mu$m) and a collimating lens (f = 250 mm) to form a collimated parallel beam. The expanded beam was blocked by an aperture and propagated a distance before incident onto the sample plane. The sample was positioned on an electric displacement stage (L505, Physik Instrumente) for axial movement, allowing it to be illuminated by probes with different wavefronts at various planes. A CMOS camera (Dhyana, TUCSEN, pixel size 6.5 $ \mu $m) is positioned several centimetres downstream of the aperture to capture the scattered intensity map of the exit wave. The motorised linear stage moved in increments of 5 mm, with the detector simultaneously acquiring a total of 11 diffraction data frames. Generally, the greater the disparity between data collected from the first and Nth planes of the sample, the higher the algorithm's convergence speed and reconstruction accuracy. This is fundamentally because data with greater variation provides richer optical field information. To induce significant changes in the wavefront of the light field incident upon the sample plane within a finite propagation distance, we introduced specific phase modulation. This modulation effect was achieved in the experiment using an aperture or lens; alternatively, diffractive optical elements such as metasurfaces could be employed. The motorised linear stage employed in this experiment has a travel range of five centimetres, which sufficiently meets the experimental requirements. Prior to running the iterative phase retrieval algorithm, a preprocessing step is required to determine the imaging distance, which can be achieved through certain autofocusing methods \cite{auto-foucs1, auto-foucs2}. Background noise is also subtracted before the data is input into the iterative algorithm.
    
    The reconstruction results are shown in Fig.~\ref{fig3}. Both the smooth contours and sharp edges in the image have been accurately restored, with artefacts completely eliminated. Due to contamination on the sample, dust particles and water stains were also clearly reconstructed, highlighting the reliability and stability of our approach. The reconstructed USAF 1951 pattern's minimum feature is the sixth group element 2, corresponding to a spatial resolution of 71.8 lines per millimetre, consistent with the sensor's maximum resolution of 76.9 lines per millimetre. The angular spectrum method describes light propagation in free space. Ideally, the reconstruction resolution depends solely on the detector pixel size of 6.5 $\mu $m. Using the Fresnel diffraction formula, the sampling interval can be made adaptive to the physical resolution. Note, however, that a wavelength-dependent scaling relationship exists for the scale of the near-field object function.
    
    Figs.~\ref{fig3}(c) and (d) show the amplitude and phase distributions of the probe function at the aperture plane, respectively. At this plane, the beam exhibits distinct boundary features due to the aperture constraint. Additionally, dust and defects (such as dead spots) on the surfaces of optical elements in the optical path interfere with the wavefront, causing the actual probe light field to deviate from the ideal plane wavefront. This effect manifests in the reconstructed probe function as corresponding non-uniform structures in both the amplitude and phase maps. Nevertheless, the method proposed in this study still achieves high-precision reconstruction of the incident light field, fully demonstrating its robustness and superior performance in wavefront detection.
    
    \subsection{spectral multiplexing experimental results}
    
    The reconstruction of wavefronts with distinct frequency components in partially coherent sources and the analysis of dispersive samples remain central research challenges within computational imaging and advanced optical metrology. Particularly in biological microscopy, biological specimens—such as cells and tissues—typically exhibit pronounced dispersive properties, with refractive indices strongly dependent upon wavelength. Whilst partially coherent light sources can provide multi-wavelength information to enhance contrast, they also introduce complex inter-frequency coupling and chromatic aberration interference. Traditional monochromatic phase reconstruction methods struggle to accurately decouple frequency-domain and spatial phase information, leading to artefacts, edge blurring, or quantisation distortion in reconstruction results. The axial ptychographic coherent diffractive imaging (AP-CDI) we propose enables simultaneous inversion of multi-wavelength wavefronts and precise characterisation of sample dispersion properties, which is crucial for achieving label-free, high-resolution, multi-parameter quantitative phase imaging (see Methods section for details).  
    
     \begin{figure*}[ht]
 	\centering\includegraphics[width=1\linewidth]{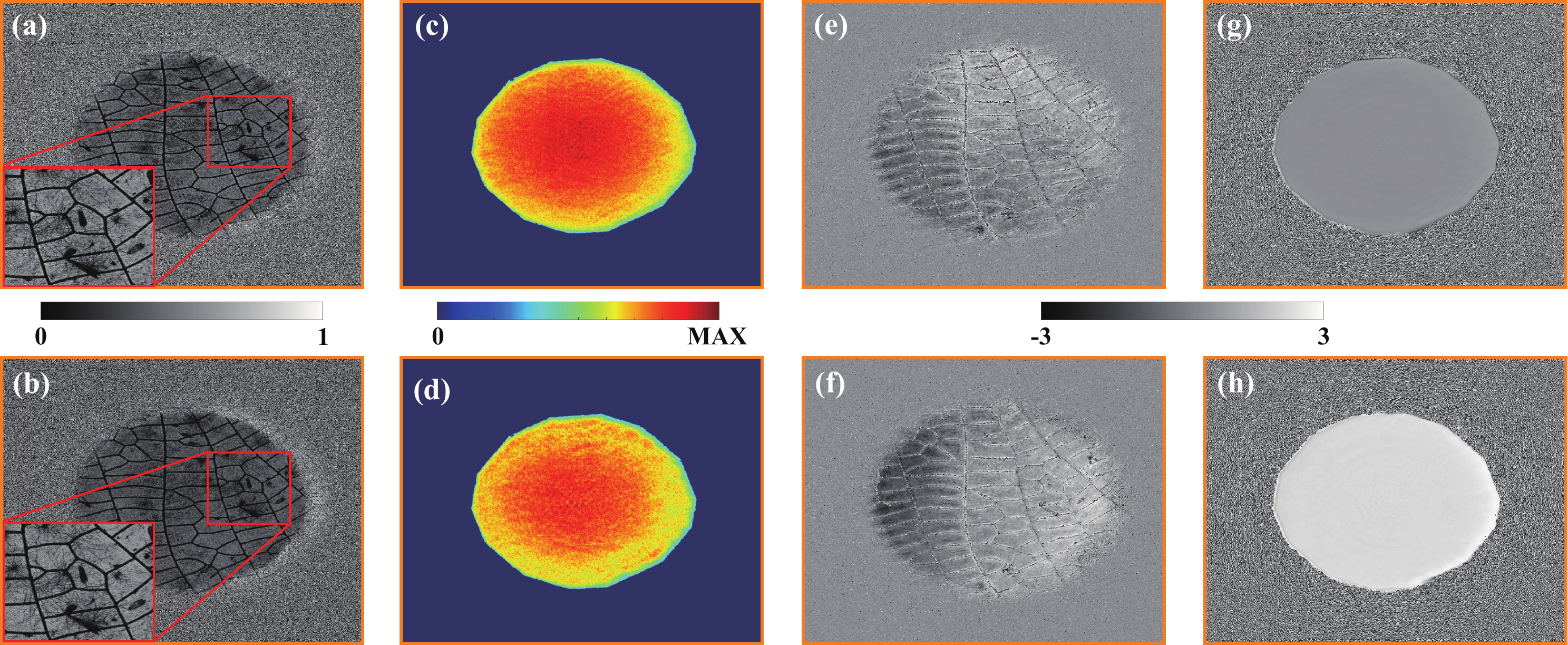}
 	\caption{ \label{fig4}\footnotesize Spectrum multiplexing experimental results of the collimated beam after passing through the aperture. (a) and (b) show the amplitude distributions of the object function reconstructed from the green and red light channels respectively; (c) and (d) display the corresponding amplitude distributions of the probe function; (e) and (f) present the phase distributions of the object function, exhibiting significant differences; (g) and (h) depict the phase distributions of the probe function. }	
     \end{figure*}   
 
    To validate the efficacy of this approach, experimental verification was conducted within the visible light spectrum. A red laser ($ \lambda$ = 632 nm) and a green laser ($ \lambda$ = 532 nm) were employed as light sources. Following beam expansion and collimation, the two beams were combined into a single coaxial optical path via a beam splitter. All other experimental configurations remained consistent with those described in the preceding section. We employed dragonfly wings as our sample, which exhibit extremely weak absorption effects in the visible light spectrum, presenting near-transparent properties. Consequently, alterations in the light field passing through the sample manifest primarily as minute phase delays. This characteristic implies that achieving effective microscopic observation and precise measurement of dragonfly wings necessitates imaging techniques highly sensitive to phase changes.
    
    The reconstruction results are shown in Fig.~\ref{fig4}, with the upper row depicting the reconstruction under green illumination ($ \lambda$ = 532 nm) and the lower row under red illumination ($ \lambda$ = 632 nm). From the intensity reconstruction images Figs.~\ref{fig4}(a) and (b), it is evident that within the visible spectrum, the absorption characteristics of the sample are relatively similar at wavelengths of 532 nm and 632 nm. However, in the phase reconstruction images Figs.~\ref{fig4}(e) and (f), a marked difference in the phase distribution exhibited by the sample at different wavelengths is clearly observable. This phenomenon may be closely related to the sample's dispersion characteristics, wherein its refractive index varies with wavelength. This further reflects distinct phase modulation behaviours across different spectral bands, potentially attributable to variations in the sample's internal structure, material composition, or optical thickness. The results demonstrate that multi-wavelength phase imaging effectively reveals optical property differences not discernible in intensity images, providing crucial evidence for distinguishing material characteristics or structural details.
      
    During the reconstruction of the sample transmission function, the algorithm presented herein can simultaneously reconstruct the probe function under different illumination modes.  Figs.~\ref{fig4}(c) and (d) respectively display the reconstructed probe amplitude distributions at the aperture plane. The gap features present in the actual aperture structure are clearly discernible, validating that this method maintains high-precision reconstruction capability for the complex amplitude distribution of the probe light field under partially coherent illumination conditions. Moreover, even when the input data volume is identical to that of monochromatic experiments, the dual-color experiment achieves high-fidelity reconstructed images. This demonstrates that our approach reduces dependence on data volume, holding promise for advancing real-time imaging research on light-matter interactions. Particularly in the field of multispectral imaging, such an efficient quantitative phase analysis scheme exhibits broad application prospects. The AP-CDI we propose accommodates a greater number of spectral channels (see Supplementary Materials for numerical simulation results).
    
    \subsection{analysis of lens dispersion}
      
    In optical system design and imaging applications, dispersion effects constitute one of the key factors constraining imaging performance and precision. The phase velocity differences across different wavelengths caused by dispersion further lead to a series of issues including axial chromatic aberration, magnification aberration, pulse broadening, and phase distortion \cite{Lens}. These severely limit the accuracy and reliability of advanced technologies such as wide-spectrum imaging and ultrafast lasers. Whilst conventional characterisation methods (such as model-based fitting analysis or interferometry) can assess dispersion properties to some extent, they remain constrained by heavy reliance on predefined models, complex system architecture, and difficulties in achieving in-situ integration. The method proposed herein enables high-precision inversion and visual characterisation of dispersion behaviour without relying on any predefined dispersion models, rendering it particularly suitable for in-situ analysis of dispersive components. Using lenses as an exemplar, the paper demonstrates the practical application of this method in characterising dispersion properties. In this experiment, a plano-convex lens with a focal length of 300 mm was employed in place of the aperture diaphragm, with all other settings remaining unchanged.
    
    \begin{figure*}[ht]
    	\centering\includegraphics[width=1\linewidth]{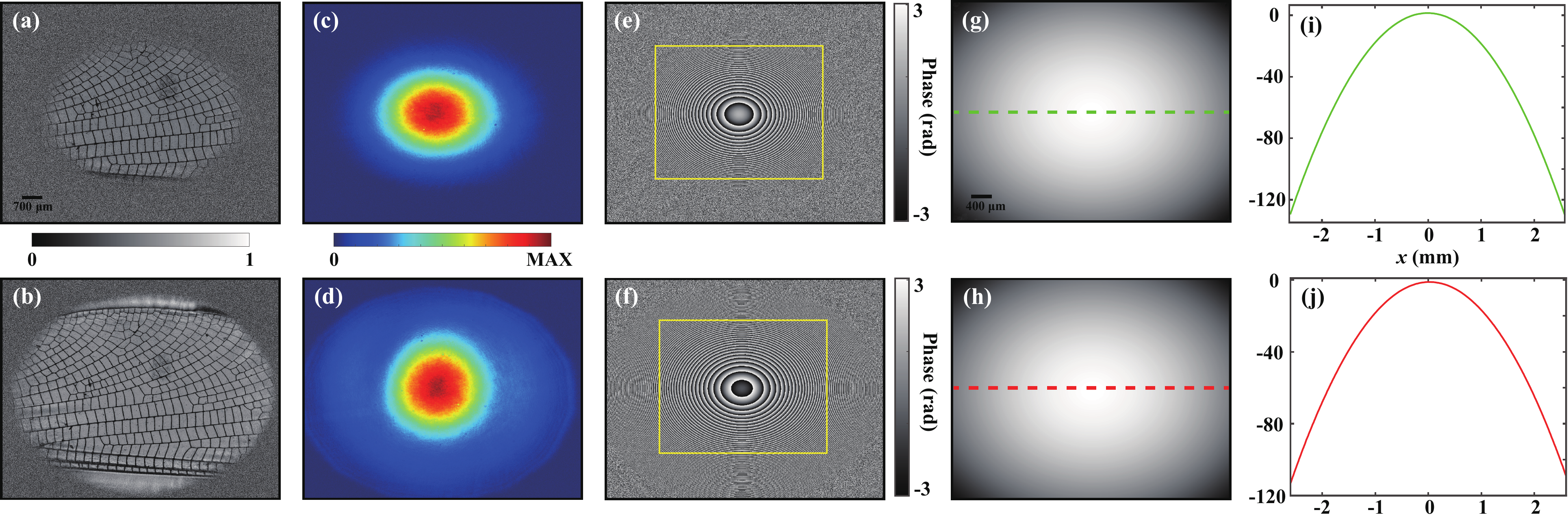}
    	\caption{\label{fig5}\footnotesize Spectrum multiplexing experimental results after the collimated beam passes through a convex lens. (a) and (b) show the reconstructed object function amplitude distributions for the green and red light channels respectively; the red light spot is larger, illuminating a greater area of the sample. (c) and (d) show the amplitude distributions of the corresponding probe functions; (e) and (f) display the probe function phases; (g) and (h) present the unwrapped phase results for the yellow regions in (e) and (f); (i) and (j) depict phase profiles along the marked lines in (g) and (h). }
    \end{figure*}
   
    Fig.~\ref{fig5}  displays reconstruction results under different illumination modes, where the top and bottom rows correspond to green light and red light illumination modes, respectively. It can be observed that the imaging field of view in the green light mode (Figure (a)) is significantly smaller than that in the red light mode (Figure (b)), primarily due to the smaller illumination spot size of green light at the sample plane. Figures (c) and (d) present the reconstructed light field amplitude distributions at the lens plane for the two modes, respectively. Spatial filtering improved the beam quality of both lasers, though the outcome depended on the initial mode content. The green laser, originating from a clean fundamental mode, produced a higher-quality Gaussian beam after filtering. The red laser's output contained significant higher-order modes; while spatial filtering attenuated these components, it could not eliminate them entirely, thereby preventing the formation of an ideal Gaussian spot. This disparity in beam purity is clearly reflected in the amplitude distribution plots. In Figures (e) and (f), we present the phase distributions of the lens at different wavelengths during reconstruction. To further analyze the phase variations of the lens, we performed envelope processing on the central regions of the phase distributions in Figures (e) and (f), yielding the results shown in Figures (g) and (h). Figures (i) and (j) display the phase profiles extracted along the dashed lines in Figures (g) and (h), respectively. Analysis of these phase profiles reveals that the phase variations under both green and red light exhibit parabolic shapes, with distinct differences between the phase profiles of the two wavelengths. The green light mode demonstrates a relatively larger amplitude of phase variation and a more pronounced focusing effect, whereas the red light mode exhibits a smaller amplitude of phase variation and a weaker focusing effect.
     
\begin{figure*}[ht]
    	\centering\includegraphics[width=1\linewidth]{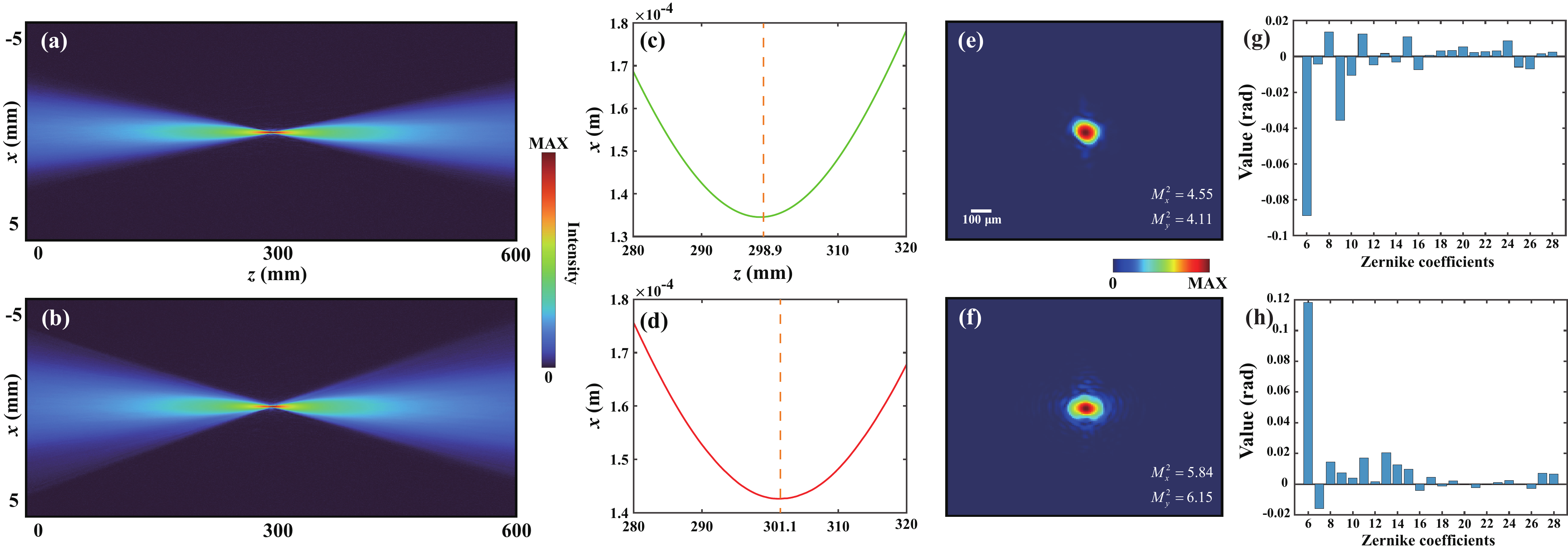}
    	\caption{\label{fig6}\footnotesize Lens Dispersion Characterisation. (a) and (b) show the logarithmic intensity distributions of the green and red light modes in the x–z plane; (c) and (d) indicate the corresponding waist positions; (e) and (f) display the beam amplitudes at the waist. Zernike polynomial fitting results of the reconstructed unwrapped phase distributions for green (g) and red (h) light. The sixth Zernike term, corresponding to astigmatism, dominates the higher-order aberrations, showing positive astigmatism for green light (elongated along y-axis) and negative astigmatism for red light (elongated along x-axis).}
    \end{figure*}

    Based on the reconstructed probe amplitude distribution, the intensity distribution of the beam at any propagation distance can be numerically calculated. Fig.~\ref{fig6} shows the reconstruction and light field propagation results for the green light (upward) and red light (downward) modes, respectively. Figures (a) and (b) display the intensity profile distributions of the two wavelength modes in the x–z plane (horizontal propagation direction) on a logarithmic scale. This method achieves wavefront reconstruction solely by translating the sample, eliminating the need to move detectors or encode/modulate the light source. Compared to traditional approaches relying on complex light source control or detector scanning, this solution maintains optical path simplicity while significantly reducing system complexity and hardware costs. It is particularly well-suited for applications demanding high real-time performance or constrained by optical configuration limitations.
    
    According to ISO~11146, the beam radius $w(z)$ is determined by calculating the second-order moment of the intensity distribution $ I_{p,z}(x, y) = \left| A_z P(x, y) \right|^2 $ of the probe light field at position $z$. For any propagation distance $z$, the beam radius along the $x$-direction, denoted as $w_x(z)$, is defined as:
    
    \begin{equation}\label{eq14}
    	w_x(z) = 2 \sqrt{\frac{\iint (x - \bar{x})^2 I_{p,z}(x, y) \, dx \, dy}{\iint I_{p,z}(x, y) \, dx \, dy}} 
    \end{equation}

    Here, $\bar{x} = \frac{\iint x I_{p,z}(x, y) dx dy}{\iint I_{p,z}(x, y) dx dy}$ denotes the $x$-coordinate of the beam's center of gravity, with the $y$-coordinate calculated similarly. As shown in Figs.~\ref{fig6}(c) and (d), the waist positions at wavelengths of 532 nm and 633 nm differ by 2.2 mm, indicating that the axial chromatic aberration of this lens is 2.2 mm. The lens material is UV fused silica, with a convex curvature radius of 137.55 mm and a nominal focal length of 300 mm at the design wavelength of 587.6 nm. The theoretical axial chromatic aberration calculated using the ideal lens model and material dispersion formula is 2.4 mm. The 0.2 mm discrepancy between the experimental and theoretical values likely stems from a combination of factors, including manufacturing tolerances of the actual lens (such as curvature radius deviation and surface profile error), the convergence accuracy of the phase reconstruction algorithm, and experimental conditions. Nevertheless, the measured results demonstrate good agreement with theoretical expectations. This not only validates that the dispersion behavior of this lens conforms to the normal dispersion characteristics of UV fused silica materials but also indicates that its manufacturing process generally meets the requirements of the optical design.
    
    The beam propagation factor $M^2$ is jointly determined by the second-order moments in both the spatial domain and the spatial frequency domain, and its expression is:    
    
    \begin{equation}\label{eq15}
    	M_x^2=4\pi \sigma_x \sigma_{kx}
    \end{equation}

    Here, $\sigma_x$ denotes the standard deviation of the light intensity distribution in the spatial domain, while $\sigma_{kx}$ represents the standard deviation of the light field in the spatial frequency domain, which can be obtained through the Fourier transform of the complex amplitude distribution. Based on the reconstructed light field complex amplitude distribution, waist spot profiles at two wavelengths are plotted as shown in  Figs.~\ref{fig6}(e) and (f). The $M^2$ factors corresponding to the green and red light modes are calculated respectively, enabling quantitative evaluation of beam quality. Quantitative analysis of beam quality indicates that the calculated $ M^2 $ factor results align with expectations. Since the green laser emits a fundamental-mode Gaussian beam with high mode quality, while the red laser operates near fundamental mode yet remains influenced by higher-order modes, the $ M^2 $ factor for green light is significantly lower than that for red light. This outcome corresponds to the distinct modal characteristics of the two lasers.
    
    The reconstructed unwrapped phase distributions (Figs.~\ref{fig5}(g) and (h)) were fitted with Zernike polynomials to quantitatively analyze the higher-order wavefront aberrations \cite{Du23,metalense1}, with the fitting results shown in Figs.~\ref{fig6}(g) and (h). The Zernike coefficients are indexed according to the standard circular pupil convention (Noll ordering), with the sixth term corresponding to astigmatism along the horizontal or vertical direction. The results show that, after removing the low-order aberrations, the sixth term dominates among the higher-order components. For the green-light phase (Figure 5(g)), the sixth term exhibits positive astigmatism, indicating a significant difference in wavefront curvature between the x and y directions, which leads to an elliptical spot elongated along the y-axis at the beam waist. In contrast, for the red-light phase (Figure 5(h)), the sixth term shows negative astigmatism, with its astigmatic axis orthogonal to that of the green light, resulting in an elliptical spot elongated along the x-axis. This analysis is fully consistent with the observed beam-waist intensity distributions, where the spots are compressed along different directions, thereby confirming that higher-order astigmatism is the primary factor governing the observed beam-spot shape variations.
    
    \subsection{spatiospectral measurement of femtosecond pulse beams}
    
    Accurate spatiotemporal characterization of broadband ultrafast laser beams is essential for achieving precise laser control \cite{felaser1,felaser2,felaser3}. Owing to the partially coherent nature of femtosecond light sources, chromatic aberrations can strongly influence the propagation and focusing behavior of the beam. Wavefront reconstruction for such partially coherent sources enables quantitative retrieval of the amplitude and phase distributions of complex optical fields, thereby providing a foundation for precise spatiotemporal control of femtosecond beams \cite{Wang}. This capability is crucial for enhancing the performance of applications such as ultrafast microscopy, coherent control, and high-precision laser–matter interaction.
    
\begin{figure*}[ht]
    	\centering\includegraphics[width=1\linewidth]{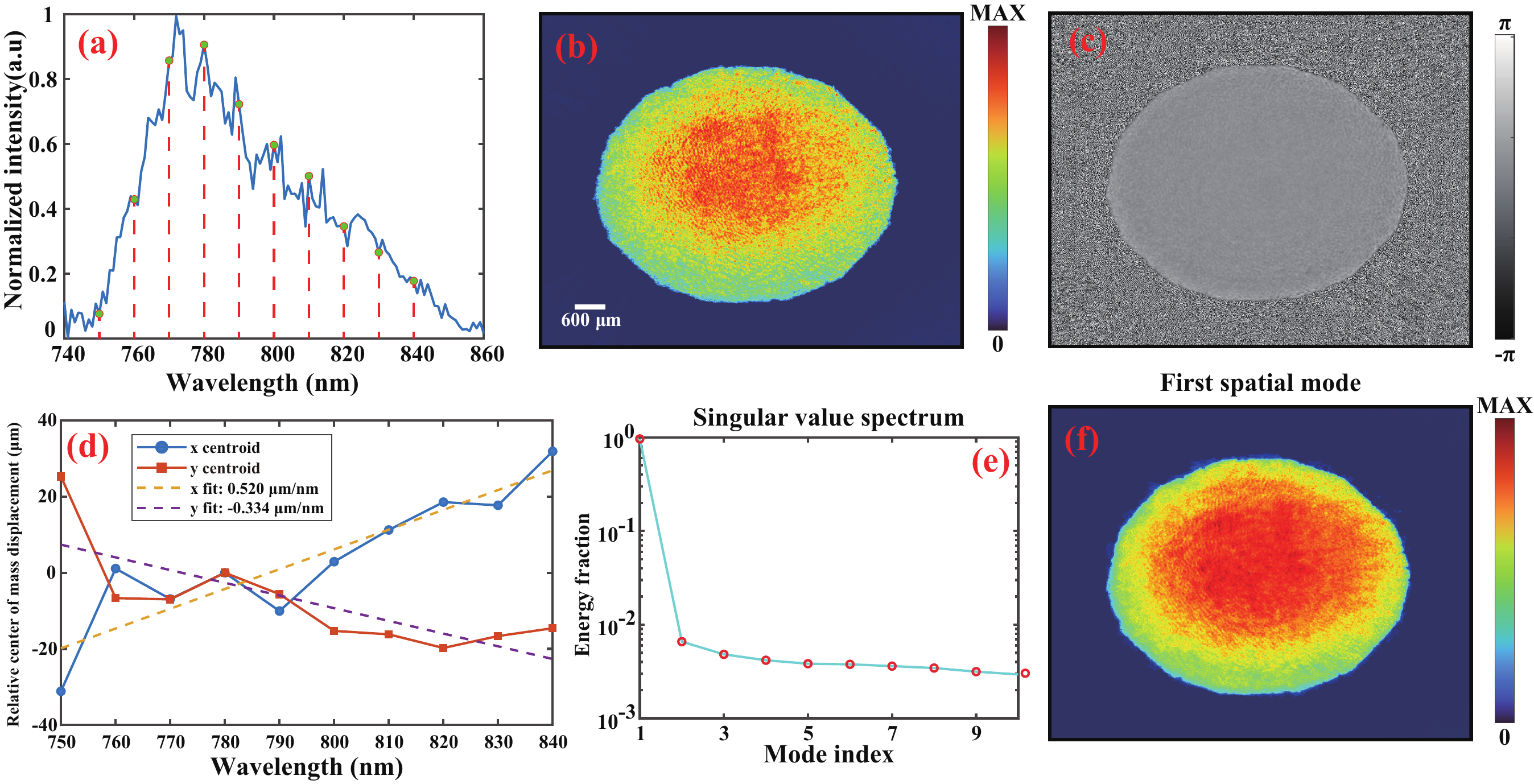}
    	\caption{\label{fig7} \footnotesize Experimental results from the femtosecond laser source. (a) Measured spectrum. (b) and (c) Reconstructed amplitude and phase of the probe function at a representative wavelength of 780 nm. (d) Relative centroid shifts of the femtosecond beam at different wavelengths, referenced to the central wavelength (780 nm), showing lateral dispersion across the spectral range. (e) Singular value decomposition (SVD) of the reconstructed complex amplitudes, with the first spatial mode dominating the energy distribution, indicating a nearly invariant spatial profile across wavelengths. (f) Amplitude distribution of the first spatial mode, representing the dominant spatial component shared across all wavelengths.}
    \end{figure*}

    We experimentally demonstrated the capability of AP-CDI for reference-free, full-field, spatiospectral characterization of ultrafast pulsed beams. The complex spatiospectral profile of a pulse beam from a Ti:sapphire amplifier (Coherent Vitara-T) was successfully reconstructed. After attenuation, the spectrum of the beam under test is shown in Fig.~\ref{fig7}(a), with a bandwidth of approximately $ \Delta \lambda / \lambda \approx 6.5\% $. The collimated beam is obstructed by an aperture and, after travelling a certain distance, strikes the sample plane. The exit wave continues to propagate towards the detector, where it is recorded as a scattering pattern. The sample is mounted on a linear translation stage and moved axially at 1 mm intervals, with data collected at 51 positions. The sample employed in this experiment consists of a dragonfly wing, primarily composed of chitin. Given that this polymeric protein compound exhibits extremely weak absorption in the infrared spectrum and virtually no spectrally specific response, this study will focus on the reconstruction results of the beam. We selected ten spectral channels with sub-bandwidths of 10 nm within the 750 nm to 840 nm range for input to the AP-CDI for reconstruction of the femtosecond laser beam. Figures (b) and (c) display the reconstruction results at the typical wavelength mode of 780 nm. Reconstruction results for typical wavelength modes are shown in Figures (b) and (c). Our approach delivers reasonable reconstruction results that accurately characterise the beam's properties. The complete results are detailed in the supplementary materials. It is noteworthy that in traditional methods, reconstructing more patterns often leads to a decline in the signal-to-noise ratio for each pattern. In contrast, our approach achieves high-quality reconstruction even under low signal-to-noise conditions without requiring data stitching, demonstrating its low dependency on data volume and excellent robustness. This characteristic is particularly advantageous for applications involving low-light conditions and demanding high sampling efficiency, such as in vivo biological imaging and rapid spectral diagnostics.
        
    To quantify the wavelength-dependent beam displacement, the centroid positions at each wavelength were calculated relative to that at the central wavelength (780 nm). As shown in Fig.~\ref{fig7}(d), the resulting relative centroid shifts reveal a lateral dispersion of the beam at the aperture plane, with differences on the order of several tens of micrometers across the spectral range. To further analyze the spatial–spectral structure of the beam, the reconstructed complex amplitudes at the aperture plane, $probe(x,y,\lambda)$, were reshaped into a two-dimensional matrix, with each column corresponding to the spatial field at a single wavelength. Singular value decomposition (SVD) was then applied to decompose the data into spatial modes and their associated spectral weights, allowing quantification of each mode’s contribution to the total energy. As shown in Fig.~\ref{fig7}(e), the first spatial mode dominates the energy distribution while higher-order modes are negligible, indicating that the spatial field profile remains nearly invariant across the spectral range, with only minor wavelength-dependent variations. Fig.~\ref{fig7}(f) displays the amplitude distribution of the first spatial mode, which is the dominant spatial component common to all wavelengths and reflects the fundamental spatial structure of the femtosecond laser beam. These analyses demonstrate that our reconstructed complex amplitudes enable a quantitative characterization of the femtosecond beam, allowing rigorous assessment of its spatial stability, coherence, and wavelength-dependent variations.
    
  \section{Discussion and Conclusions}

   In this study, we propose and experimentally validate a spectral-multiplexing lensless quantitative phase imaging framework that enables spatial–spectral wavefront characterization under partially coherent illumination. By combining multi-wavelength illumination with axial scanning, the method achieves efficient data acquisition and high-fidelity reconstruction of both the sample’s complex transmittance and the spectrally varying probe fields. The system requires no imaging lenses, reference beam, spectral filters, or interferometers, offering a compact and low-cost solution for multispectral quantitative imaging.
   
   To demonstrate its capability for in-situ characterization of dispersive optical components, the proposed method was first applied to lens analysis under discrete multi-wavelength illumination. Experimental reconstructions of the transmitted amplitude and phase distributions at 532 nm and 633 nm enabled quantitative extraction of the beam waist positions and $M^2$ factors. The measured axial chromatic shift of 2.2 mm between the two wavelengths agrees well with theoretical predictions, confirming the accuracy of this approach in assessing chromatic dispersion and wavefront aberrations.
   
   Furthermore, the proposed framework is readily extendable to continuous broadband spectral analysis. We applied it to the characterization of a Ti:sapphire femtosecond amplifier beam, successfully reconstructing spatial–spectral amplitude and phase profiles across multiple wavelength channels within the broadband emission. The retrieved results revealed the dispersion behavior and modal variations across wavelengths, demonstrating the method’s ability to quantitatively characterize partially coherent ultrafast sources. Expanding the reconstruction to a greater number of spectral modes can be achieved by employing finer axial sampling and an extended scanning range to increase data redundancy and capture the diffraction diversity of individual modes.
   
   Algorithmically, the iterative spectral–spatial reconstruction is implemented through fully vectorized GPU operations, ensuring high computational efficiency. On an NVIDIA RTX 4060 GPU (8 GB VRAM), the reconstruction for a 1024 $\times$ 1024 $\times$ 11 pixel dataset in dual-wavelength mode typically yields satisfactory results within approximately ten seconds. Moreover, the method requires only a small amount of input data to achieve stable and accurate reconstruction, which substantially reduces data acquisition demands and further enhances its feasibility for real-time and high-throughput imaging applications.
   
   In summary, the proposed spectral-multiplexing lensless imaging technique provides a compact, low-cost, and computationally scalable solution for quantitative multispectral wavefront analysis. Its compatibility with both discrete and broadband light sources demonstrates broad applicability across diverse illumination conditions—from microscopic imaging to optical metrology and beam diagnostics. By combining simplified hardware architecture, minimal data requirements, and scalable algorithmic reconstruction, this approach establishes a versatile and efficient platform for advancing multispectral quantitative imaging and wavefront characterization of partially coherent sources.
   
 \section{Methods}
 
 \subsection{axial ptychographic coherent diffractive imaging}
 
    In this section, we provide a detailed description of the axial ptychographic coherent diffractive imaging (AP-CDI). The required input data consists of a set of scattering images collected from the samples at different distances. Although the absolute axial positions of the samples and the detector (such as the closest point or the farthest point) are determined through prior measurement or through a single-step reverse propagation of the diffraction pattern, the relative distances between all other collected planes are precisely known through the controlled movement of the motorized stage. In the experiment, we collected data by moving the sample at equal intervals; however, in each iteration of the algorithm, we randomly sampled data from different planes for computation, which effectively accelerated the convergence rate.
    
    Typically, we use the single-step backpropagation results from the probe scattering plot as the initial estimate for the probe function, and initialize the object function with a random distribution. At iteration $j$, for the exit wave at the $n$th acquisition plane:
    
    \begin{equation}\label{eq16}
    	\psi_{j,n,\lambda_i}(\mathbf{r})=P_{j,n,\lambda_i}(\mathbf{r}) \cdot O_{j,\lambda_i}(\mathbf{r}) 
    \end{equation}
 
     The light field at the detector plane is recorded as:
     
     \begin{equation}\label{eq17}
     	\varPsi_{j,n,\lambda_i}(\mathbf{R})=A_{\Delta z_n,\lambda_i} \psi_{j,n,\lambda_i}(\mathbf{r})
     \end{equation}
      
     Here, $\mathbf{R} = (X, Y)$ denotes the two-dimensional coordinate system of the detector plane. In practice, since the detector's integration time far exceeds the coherence time between different wavelengths, the recorded incoherent diffraction pattern can be regarded as a spectrally weighted, incoherent superposition of the diffraction patterns corresponding to each wavelength component. Based on Wolf's coherent mode theory \cite{Wolf1982}, a partially coherent wave can be decomposed into a set of orthogonal coherent modes propagating independently in free space. This theory provides the theoretical foundation for reconstructing multiple modes in partially coherent fields. When the source is incoherent, the contribution of all wavelength components to the diffraction pattern must be considered, specifically expressed as:
     
      \begin{equation}\label{eq18}
     	I_{j,n}=\sum_{i}\left| 	\varPsi_{j,n,\lambda_i}(\mathbf{R}) \right|^{2}
      \end{equation}	
     
     As described in Section 2A, impose  modulus constraints on the far-field light field at different wavelengths and update the exit wave function:
     
\begin{equation}\label{eq19}
	\begin{aligned}
		\varPsi_{j,n,\lambda_i}^{'}(\mathbf{R}) &= \varPsi_{j,n,\lambda_i}(\mathbf{R}) \sqrt{\frac{I_{n}^{m}}{I_{j,n}}} \\
		\psi_{j,n,\lambda_i}^{'}(\mathbf{r}) &= A_{\Delta -z_n,\lambda_i} \varPsi_{j,n,\lambda_i}^{'}(\mathbf{R})
	\end{aligned}
\end{equation}
 
     The intensity measurement $ I_{n}^{m} $ corresponds to the diffraction pattern captured when the sample is positioned at the $n$th location. The constraint in Eq.~\ref{eq19} ensures that the incoherent sum of the far-field intensities from all wavelength modes equals the measured data $ I_{n}^{m} $. Regularization constraints are incorporated into the updates of the object and probe functions:
     
 \begin{equation}\label{eq20}
 	\begin{aligned}
 		O^{'}_{j,\lambda_i}(\mathbf{r}) &= O_{j,\lambda_i}(\mathbf{r}) +   \frac{ P_{j,n,\lambda_i}^*(\mathbf{\mathbf{r}}) \cdot [  \psi_{j,n,\lambda_i}^{'}-\psi_{j,n,\lambda_i}(\mathbf{r}) ] }{ \alpha \left| P_{j,n,\lambda_i}(\mathbf{r}) \right|_ {\operatorname{ max}}^2 + (1-\alpha) \left|   P_{j,n,\lambda_i}(\mathbf{r}) \right|^2}  \\
 		P^{'}_{j,n,\lambda_i}(\mathbf{r}) &= P_{j,n,\lambda_i}(\mathbf{r}) +   \frac{ O_{j,\lambda_i}^*(\mathbf{\mathbf{r}}) \cdot [  \psi_{j,n,\lambda_i}^{'}-\psi_{j,n,\lambda_i}(\mathbf{r}) ] }{ \beta \left| O_{j,\lambda_i}(\mathbf{r}) \right|_ {\operatorname{ max}}^2 + (1-\beta) \left|   O_{j,\lambda_i}(\mathbf{r}) \right|^2} 
 	\end{aligned}
 \end{equation}

      $\alpha $ and $ \beta $ are constants between 0 and 1.Before the sample is placed in the optical path, the diffraction pattern $ I_{p}^{m} $ of the direct beam collected at the detector is used for subsequent probe modulus constraint. The updated probe is propagated to the detector plane, forming the estimated value of the probe on the detector, given by $ \mathscr{P}_{j,n,\lambda_i}(\mathbf{R}) = A_{\Delta z_n,\lambda_i} P^{'}_{j,n,\lambda_i}(\mathbf{r})$. Similar to Eq.~\ref{eq19}, we impose a modulus constraint on the probe's estimated value to ensure the accuracy of the probe's measurement results:
      
     \begin{equation}\label{eq21}
     	\begin{aligned}
     	    I_{j,n,p} &=\sum_{i}\left| \mathscr{P}_{j,n,\lambda_i}(\mathbf{R}) \right|^{2} \\
            \mathscr{P}_{j,n,\lambda_i}^{'}(\mathbf{R}) &= \mathscr{P}_{j,n,\lambda_i}(\mathbf{R}) \sqrt{\frac{I_{p}^{m}}{I_{j,n,p}}}
     	\end{aligned} 		 
     \end{equation}    
      
    This modulus-constrained probe guess is backpropagated to the sample plane, forming a further updated probe estimate:
    $ P_{j,n,\lambda_i}^{''}(\mathbf{r}) = A_{\Delta -z_n,\lambda_i} \mathscr{P}_{j,n,\lambda_i}^{'}(\mathbf{R}) $. The additional constraint imposed by the probe further narrows the range of possible probe values, thereby providing more information and accelerating convergence \cite{13.5nm}. 
      
      After performing the above operation on the diffraction data for all N planes, one iteration is completed.
      
	\begin{backmatter}
		\bmsection{Funding} National Key Research and Development Program of China (2024YFE0205800); National Natural Science Foundation of China (12274158, 12021004).
		
		\bmsection{Disclosures} The authors declare no conflicts of interest.
		
		\bmsection{Data availability} Data underlying the results are not publicly available at this time but may be obtained from the authors upon reasonable request.
		
		
	\end{backmatter}
	
	\bibliography{bibliography}

@article{Aidukas2024,
	author = {Aidukas, Tomas and Phillips, Nicholas W. and Diaz, Ana and Poghosyan, Emiliya and M{\"u}ller, Elisabeth and Levi, A. F. J. and Aeppli, Gabriel and Guizar-Sicairos, Manuel and Holler, Mirko},
	title = {High-performance 4-nm-resolution X-ray tomography using burst ptychography},
	journal = {Nature},
	year = {2024},
	volume = {632},
	number = {8023},
	pages = {81--88},
	issn = {1476-4687},
	doi = {10.1038/s41586-024-07615-6},
	url = {https://doi.org/10.1038/s41586-024-07615-6},
	month = {08},
	date = {2024/08/01}
}

@article{Daurer:21,
	author = {Benedikt J. Daurer and Simone Sala and Max F. Hantke and Hemanth K. N. Reddy and Johan Bielecki and Zhou Shen and Carl Nettelblad and Martin Svenda and Tomas Ekeberg and Gabriella A. Carini and Philip Hart and Timur Osipov and Andrew Aquila and N. Duane Loh and Filipe R. N. C. Maia and Pierre Thibault},
	journal = {Optica},
	number = {4},
	pages = {551--562},
	publisher = {Optica Publishing Group},
	title = {Ptychographic wavefront characterization for single-particle imaging at x-ray lasers},
	volume = {8},
	month = {Apr},
	year = {2021},
	url = {https://opg.optica.org/optica/abstract.cfm?URI=optica-8-4-551},
	doi = {10.1364/OPTICA.416655},
}

@article{ph2,
	author = {J. Miao and D. Sayre and H. N. Chapman},
	journal = {J. Opt. Soc. Am. A},
	number = {6},
	pages = {1662--1669},
	publisher = {Optica Publishing Group},
	title = {Phase retrieval from the magnitude of the Fourier transforms of nonperiodic objects},
	volume = {15},
	month = {Jun},
	year = {1998},
	url = {https://opg.optica.org/josaa/abstract.cfm?URI=josaa-15-6-1662},
	doi = {10.1364/JOSAA.15.001662},
}

@article{ph3,
	author = {J. R. Fienup},
	journal = {Appl. Opt.},
	number = {15},
	pages = {2758--2769},
	publisher = {Optica Publishing Group},
	title = {Phase retrieval algorithms: a comparison},
	volume = {21},
	month = {Aug},
	year = {1982},
	url = {https://opg.optica.org/ao/abstract.cfm?URI=ao-21-15-2758},
	doi = {10.1364/AO.21.002758},
}

@article{ph7,
	author = {Zhang, Fucai and Chen, Bo and Morrison, Graeme R. and Vila-Comamala, Joan and Guizar-Sicairos, Manuel and Robinson, Ian K.},
	title = {Phase retrieval by coherent modulation imaging},
	journal = {Nature Communications},
	volume = {7},
	number = {1},
	pages = {13367},
	ISSN = {2041-1723},
	DOI = {10.1038/ncomms13367},
	url = {https://doi.org/10.1038/ncomms13367},
	year = {2016},
	type = {Journal Article}
}

@ARTICLE{ph8,
	author={Shechtman, Yoav and Eldar, Yonina C. and Cohen, Oren and Chapman, Henry Nicholas and Miao, Jianwei and Segev, Mordechai},
	journal={IEEE Signal Processing Magazine}, 
	title={Phase Retrieval with Application to Optical Imaging: A contemporary overview}, 
	year={2015},
	volume={32},
	number={3},
	pages={87-109},
	doi={10.1109/MSP.2014.2352673}}

@article{bio1,
	author = {Liu, Chang and Eschen, Wilhelm and Loetgering, Lars and {Penagos Molina}, Daniel S. and Klas, Robert and Iliou, Alexander and Steinert, Michael and Herkersdorf, Sebastian and Kirsche, Alexander and Pertsch, Thomas and Hillmann, Falk and Limpert, Jens and Rothhardt, Jan},
	title = {Visualizing the Ultra-Structure of Microorganisms Using Table-Top Extreme Ultraviolet Imaging},
	journal = {PhotoniX},
	year = {2023},
	volume = {4},
	number = {1},
	pages = {6},
	issn = {2662-1991},
	doi = {10.1186/s43074-023-00084-6},
	url = {https://doi.org/10.1186/s43074-023-00084-6},
	month = {01},
	date = {2023/01/24}
}

@article{bio2,
	author = {Jones, Michael W. M. and Elgass, Kirstin and Junker, Mark D. and Luu, Mac B. and Ryan, Michael T. and Peele, Andrew G. and {van Riessen}, Grant A.},
	title = {Mapping Biological Composition Through Quantitative Phase and Absorption X-ray Ptychography},
	journal = {Scientific Reports},
	year = {2014},
	volume = {4},
	number = {1},
	pages = {6796},
	issn = {2045-2322},
	doi = {10.1038/srep06796},
	url = {https://doi.org/10.1038/srep06796},
	month = {10},
	date = {2014/10/28}
}

@article{ptychography1,
	title = {A Phase Retrieval Algorithm for Shifting Illumination},
	author = {Rodenburg, J. M. and Faulkner, Helen Mary Louise},
	year = {2004},
	journal = {Applied Physics Letters},
	volume = {85},
	number = {20},
	pages = {4795--4797},
	issn = {0003-6951},
	publisher = {AIP Publishing},
	language = {English}
}

@article{ptychography2,
	title = {Movable Aperture Lensless Transmission Microscopy: A Novel Phase Retrieval Algorithm},
	author = {Faulkner, Helen Mary Louise and Rodenburg, J. M.},
	year = {2004},
	journal = {Physical Review Letters},
	volume = {93},
	number = {2},
	pages = {023903-1--023903-4},
	issn = {0031-9007},
	publisher = {American Physical Society},
	language = {English}
}

@article{broadband2,
	author = {Thibault, Pierre and Menzel, Andreas},
	title = {Reconstructing state mixtures from diffraction measurements},
	journal = {Nature},
	volume = {494},
	number = {7435},
	pages = {68-71},
	ISSN = {1476-4687},
	DOI = {10.1038/nature11806},
	url = {https://doi.org/10.1038/nature11806},
	year = {2013},
	type = {Journal Article}
}

@article{broadband3,
	title = {Information multiplexing in ptychography},
	journal = {Ultramicroscopy},
	volume = {138},
	pages = {13-21},
	year = {2014},
	issn = {0304-3991},
	doi = {https://doi.org/10.1016/j.ultramic.2013.12.003},
	url = {https://www.sciencedirect.com/science/article/pii/S030439911300315X},
	author = {Darren J. Batey and Daniel Claus and John M. Rodenburg},
}

@article{broadband5,
	author = {Witte, Stefan and Tenner, Vasco T. and Noom, Daniel W. E. and Eikema, Kjeld S. E.},
	title = {Lensless diffractive imaging with ultra-broadband table-top sources: from infrared to extreme-ultraviolet wavelengths},
	journal = {Light: Science \& Applications},
	volume = {3},
	number = {3},
	pages = {e163-e163},
	ISSN = {2047-7538},
	DOI = {10.1038/lsa.2014.44},
	url = {https://doi.org/10.1038/lsa.2014.44},
	year = {2014},
	type = {Journal Article}
}

@article{broadband7,
	title = {Potential of Attosecond Coherent Diffractive Imaging},
	author = {Rana, Arjun and Zhang, Jianhua and Pham, Minh and Yuan, Andrew and Lo, Yuan Hung and Jiang, Huaidong and Osher, Stanley J. and Miao, Jianwei},
	journal = {Phys. Rev. Lett.},
	volume = {125},
	issue = {8},
	pages = {086101},
	numpages = {6},
	year = {2020},
	month = {Aug},
	publisher = {American Physical Society},
	doi = {10.1103/PhysRevLett.125.086101},
	url = {https://link.aps.org/doi/10.1103/PhysRevLett.125.086101}
}

@article{broadband8,
	title = {Broadband ptychographic imaging with an accurately sampled spectrum},
	author = {Liu, Ruifan and Cao, Wei and You, Qijun and Gao, Yun and Zhang, Fucai and Liao, Qing and Lu, Peixiang},
	journal = {Phys. Rev. A},
	volume = {107},
	issue = {3},
	pages = {033510},
	numpages = {9},
	year = {2023},
	month = {Mar},
	publisher = {American Physical Society},
	doi = {10.1103/PhysRevA.107.033510},
	url = {https://link.aps.org/doi/10.1103/PhysRevA.107.033510}
}

@article{broadband10,
	title = {Interface and spectrum multiplexing ptychographic reflection microscopy},
	journal = {Optics and Lasers in Engineering},
	volume = {193},
	pages = {109076},
	year = {2025},
	issn = {0143-8166},
	doi = {https://doi.org/10.1016/j.optlaseng.2025.109076},
	url = {https://www.sciencedirect.com/science/article/pii/S0143816625002623},
	author = {Yun Gao and Qijun You and Peixiang Lu and Wei Cao}
}

@article{You,
	author = {Qijun You and Lingshuo Meng and Yun Gao and Qing Liao and Wei Cao and Peixiang Lu},
	journal = {Opt. Lett.},
	number = {19},
	pages = {5535--5538},
	publisher = {Optica Publishing Group},
	title = {Spectral multiplexing based on multi-distance lensless imaging},
	volume = {49},
	month = {Oct},
	year = {2024},
	url = {https://opg.optica.org/ol/abstract.cfm?URI=ol-49-19-5535},
	doi = {10.1364/OL.534585},
}

@article{Du23,
	author = {Mengqi Du and Xiaomeng Liu and Antonios Pelekanidis and Fengling Zhang and Lars Loetgering and Patrick Konold and Christina L. Porter and Peter Smorenburg and Kjeld S. E. Eikema and Stefan Witte},
	journal = {Optica},
	number = {2},
	pages = {255--263},
	publisher = {Optica Publishing Group},
	title = {High-resolution wavefront sensing and aberration analysis of multi-spectral extreme ultraviolet beams},
	volume = {10},
	month = {Feb},
	year = {2023},
	url = {https://opg.optica.org/optica/abstract.cfm?URI=optica-10-2-255},
	doi = {10.1364/OPTICA.478346},	
}

@article{Md,
	author = {Giancarlo, Pedrini and Wolfgang, Osten and Yan, Zhang},
	title = {Wave-front reconstruction from a sequence of interferograms recorded at different planes},
	journal = {Opt. Lett.},
	volume = {30},
	number = {8},
	pages = {833--835},
	DOI = {10.1364/OL.30.000833},
	url = {https://opg.optica.org/ol/abstract.cfm?URI=ol-30-8-833},
	year = {2005},
	type = {Journal Article}
}

@article{Md2,
	author = {Guo, Cheng and Shen, Cheng and Li, Qiang and Tan, Jiubin and Liu, Shutian and Kan, Xinchi and Liu, Zhengjun},
	title = {A fast-converging iterative method based on weighted feedback for multi-distance phase retrieval},
	journal = {Scientific Reports},
	volume = {8},
	number = {1},
	pages = {6436},
	ISSN = {2045-2322},
	DOI = {10.1038/s41598-018-24666-8},
	url = {https://doi.org/10.1038/s41598-018-24666-8},
	year = {2018},
	type = {Journal Article}
}

@article{metalense1,
	title = {Metalenses Phase Characterization by Multi-Distance Phase Retrieval},
	author = {Liu, Bowen and Cheng, Jialuo and Zhao, Maoxiong and Yao, Jin and Liu, Xiaoyuan and Chen, Shaohu and Shi, Lei and Tsai, Din Ping and Geng, Zihan and Chen, Mu Ku},
	year = {2024},
	month = {08},
	date = {2024-08-06},
	journal = {Light: Science \& Applications},
	volume = {13},
	number = {1},
	pages = {182},
	issn = {2047-7538},
	doi = {10.1038/s41377-024-01530-1},
	url = {https://doi.org/10.1038/s41377-024-01530-1},
	publisher = {Springer Nature},
	language = {en}
}

@article{metalense2,
	author = {Cheng, Jialuo and Zhou, Yin and Gao, Yunhui and Liu, Bowen and Liu, Xiaoyuan and Shi, Lei and Geng, Zihan and Chen, Mu Ku},
	title = {Quantitative Phase Imaging for Meta-Lenses by Phase Retrieval},
	journal = {Advanced Optical Materials},
	volume = {13},
	number = {9},
	pages = {2402833},
	doi = {https://doi.org/10.1002/adom.202402833},
	url = {https://advanced.onlinelibrary.wiley.com/doi/abs/10.1002/adom.202402833},
	eprint = {https://advanced.onlinelibrary.wiley.com/doi/pdf/10.1002/adom.202402833},
	year = {2025}
}

@article{metalense3,
	author = {Chuanjian Zheng and Wenli Wang and Yanfang Ji and Yao Hu and Shaohui Zhang and Qun Hao},
	journal = {Optica},
	number = {3},
	pages = {374--383},
	publisher = {Optica Publishing Group},
	title = {High-space--bandwidth product characterization of metalenses by information fusion of multi-angle illumination},
	volume = {12},
	month = {Mar},
	year = {2025},
	url = {https://opg.optica.org/optica/abstract.cfm?URI=optica-12-3-374},
	doi = {10.1364/OPTICA.551944},
}

@article{metalense4,
	title = {Quantitative Phase Imaging Endoscopy with a Metalens},
	author = {Shanker, Aamod and Fr{\"o}ch, Johannes E. and Mukherjee, Saswata and Zhelyeznyakov, Maksym and Brunton, Steven L. and Seibel, Eric J. and Majumdar, Arka},
	year = {2024},
	month = {11},
	date = {2024-11-08},
	journal = {Light: Science \& Applications},
	volume = {13},
	number = {1},
	pages = {305},
	issn = {2047-7538},
	doi = {10.1038/s41377-024-01587-y},
	url = {https://doi.org/10.1038/s41377-024-01587-y},
	publisher = {Springer Nature},
	language = {en}
}

@article{encode1,
	title = {Wide-Field, High-Resolution Fourier Ptychographic Microscopy},
	author = {Zheng, Guoan and Horstmeyer, Roarke and Yang, Changhuei},
	year = {2013},
	month = {09},
	date = {2013-09-01},
	journal = {Nature Photonics},
	volume = {7},
	number = {9},
	pages = {739--745},
	issn = {1749-4893},
	doi = {10.1038/nphoton.2013.187},
	url = {https://doi.org/10.1038/nphoton.2013.187},
	publisher = {Nature Publishing Group},
	language = {en}
}

@ARTICLE{encode2,
	author={Boominathan, Vivek and Adams, Jesse K. and Robinson, Jacob T. and Veeraraghavan, Ashok},
	journal={IEEE Transactions on Pattern Analysis and Machine Intelligence}, 
	title={PhlatCam: Designed Phase-Mask Based Thin Lensless Camera}, 
	year={2020},
	volume={42},
	number={7},
	pages={1618-1629},
	doi={10.1109/TPAMI.2020.2987489}
}

@article{encode3,
	title = {Fast, long-term, super-resolution imaging with Hessian structured illumination microscopy},
	author = {Huang, Xiaojie and Fan, Jiji and Li, Liang and Liu, Hongtao and Wu, Ruxin and Wu, Yicong and Wei, Lai and Mao, Haoyu and Lal, Amit and Xi, Peng and others},
	year = {2018},
	journal = {Nature Biotechnology},
	volume = {36},
	number = {5},
	pages = {451--459},
	doi = {10.1038/nbt.4115}
}

@article{CCTV1,
	title = {Iterative Projection Meets Sparsity Regularization: Towards Practical Single-Shot Quantitative Phase Imaging with In-Line Holography},
	author = {Gao, Yunhui and Cao, Liangcai},
	journal = {Light: Advanced Manufacturing},
	year = {2023},
	volume = {4},
	number = {LAM2022050014},
	pages = {37},
	issn = {2689-9620},
	doi = {10.37188/lam.2023.006},
	url = {https://www.light-am.com/article/id/e708f1fe-26f9-40d8-a314-13f06f490ae0}
}

@article{CCTV2,
	title = {Twin-Image-Free Holography: A Compressive Sensing Approach},
	author = {Zhang, Wenhui and Cao, Liangcai and Brady, David J. and Zhang, Hua and Cang, Ji and Zhang, Hao and Jin, Guofan},
	journal = {Phys. Rev. Lett.},
	volume = {121},
	issue = {9},
	pages = {093902},
	numpages = {6},
	year = {2018},
	month = {Aug},
	publisher = {American Physical Society},
	doi = {10.1103/PhysRevLett.121.093902},
	url = {https://link.aps.org/doi/10.1103/PhysRevLett.121.093902}
}

@inproceedings{Fourier,
	title={Introduction to Fourier optics},
	author={Joseph W. Goodman},
	year={1969},
	url={https://api.semanticscholar.org/CorpusID:118908270}
}

@ARTICLE{error1,
	author={Candès, Emmanuel J. and Li, Xiaodong and Soltanolkotabi, Mahdi},
	journal={IEEE Transactions on Information Theory}, 
	title={Phase Retrieval via Wirtinger Flow: Theory and Algorithms}, 
	year={2015},
	volume={61},
	number={4},
	pages={1985-2007},
	doi={10.1109/TIT.2015.2399924}
}

@ARTICLE{error2,
	author={Wang, Gang and Giannakis, Georgios B. and Eldar, Yonina C.},
	journal={IEEE Transactions on Information Theory}, 
	title={Solving Systems of Random Quadratic Equations via Truncated Amplitude Flow}, 
	year={2018},
	volume={64},
	number={2},
	pages={773-794},
	doi={10.1109/TIT.2017.2756858}
}

@article{epie,
	title = {An improved ptychographical phase retrieval algorithm for diffractive imaging},
	journal = {Ultramicroscopy},
	volume = {109},
	number = {10},
	pages = {1256-1262},
	year = {2009},
	issn = {0304-3991},
	doi = {https://doi.org/10.1016/j.ultramic.2009.05.012},
	url = {https://www.sciencedirect.com/science/article/pii/S0304399109001284},
	author = {Andrew M. Maiden and John M. Rodenburg},
}

@article{rpie,
	author = {Andrew Maiden and Daniel Johnson and Peng Li},
	journal = {Optica},
	number = {7},
	pages = {736--745},
	publisher = {Optica Publishing Group},
	title = {Further improvements to the ptychographical iterative engine},
	volume = {4},
	month = {Jul},
	year = {2017},
	url = {https://opg.optica.org/optica/abstract.cfm?URI=optica-4-7-736},
	doi = {10.1364/OPTICA.4.000736},
}

@article{auto-foucs1,
	title = {Lensfree auto-focusing imaging using nuclear norm of gradient},
	journal = {Optics and Lasers in Engineering},
	volume = {156},
	pages = {107076},
	year = {2022},
	issn = {0143-8166},
	doi = {https://doi.org/10.1016/j.optlaseng.2022.107076},
	url = {https://www.sciencedirect.com/science/article/pii/S0143816622001282},
	author = {Cheng Guo and Feilong Zhang and Xianming Liu and Qiang Li and Shenghao Zheng and Jiubin Tan and Zhengjun Liu and Weibo Wang},
}

@article{auto-foucs2,
	title = {Fast automatic multiple positioning for lensless coherent diffraction imaging},
	journal = {Optics and Lasers in Engineering},
	volume = {155},
	pages = {107055},
	year = {2022},
	issn = {0143-8166},
	doi = {https://doi.org/10.1016/j.optlaseng.2022.107055},
	url = {https://www.sciencedirect.com/science/article/pii/S0143816622001105},
	author = {Xuyang Zhou and Xiu Wen and Yu Ji and Yong Geng and Shutian Liu and Zhengjun Liu},

}

@article{Lens,
	author = {Z. Bor},
	journal = {Opt. Lett.},
	number = {2},
	pages = {119--121},
	publisher = {Optica Publishing Group},
	title = {Distortion of femtosecond laser pulses in lenses},
	volume = {14},
	month = {Jan},
	year = {1989},
	url = {https://opg.optica.org/ol/abstract.cfm?URI=ol-14-2-119},
	doi = {10.1364/OL.14.000119},
}

@ARTICLE{felaser1,
	author={Dorrer, Christophe},
	journal={IEEE Journal of Selected Topics in Quantum Electronics}, 
	title={Spatiotemporal Metrology of Broadband Optical Pulses}, 
	year={2019},
	volume={25},
	number={4},
	pages={1-16},
	doi={10.1109/JSTQE.2019.2899019}
}

@article{felaser2,
	author = {Hampus Wikmark  and Chen Guo  and Jan Vogelsang  and Peter W. Smorenburg  and Hélène Coudert-Alteirac  and Jan Lahl  and Jasper Peschel  and Piotr Rudawski  and Hugo Dacasa  and Stefanos Carlström  and Sylvain Maclot  and Mette B. Gaarde  and Per Johnsson  and Cord L. Arnold  and Anne L’Huillier },
	title = {Spatiotemporal coupling of attosecond pulses},
	journal = {Proceedings of the National Academy of Sciences},
	volume = {116},
	number = {11},
	pages = {4779-4787},
	year = {2019},
	doi = {10.1073/pnas.1817626116},
	URL = {https://www.pnas.org/doi/abs/10.1073/pnas.1817626116},
	eprint = {https://www.pnas.org/doi/pdf/10.1073/pnas.1817626116},
}

@article{felaser3,
	doi = {10.1088/2040-8986/abad08},
	url = {https://doi.org/10.1088/2040-8986/abad08},
	year = {2020},
	month = {sep},
	publisher = {IOP Publishing},
	volume = {22},
	number = {10},
	pages = {103501},
	author = {Jolly, Spencer W and Gobert, Olivier and Quéré, Fabien},
	title = {Spatio-temporal characterization of ultrashort laser beams: a tutorial},
	journal = {Journal of Optics},
}

@article{Wolf1982,
	author = {Emil Wolf},
	journal = {J. Opt. Soc. Am.},
	number = {3},
	pages = {343--351},
	publisher = {Optica Publishing Group},
	title = {New theory of partial coherence in the space--frequency domain. Part I: spectra and cross spectra of steady-state sources},
	volume = {72},
	month = {Mar},
	year = {1982},
	url = {https://opg.optica.org/abstract.cfm?URI=josa-72-3-343},
	doi = {10.1364/JOSA.72.000343},
}

@article{13.5nm,
	title = {Subwavelength Coherent Imaging of Periodic Samples Using a 13.5 nm Tabletop High-Harmonic Light Source},
	author = {Gardner, Dennis F. and Tanksalvala, Michael and Shanblatt, Elisabeth R. and Zhang, Xiaoshi and Galloway, Benjamin R. and Porter, Christina L. and {Karl Jr}, Robert and Bevis, Charles and Adams, Daniel E. and Kapteyn, Henry C. and Murnane, Margaret M. and Mancini, Giulia F.},
	year = {2017},
	month = {04},
	date = {2017-04-01},
	journal = {Nature Photonics},
	volume = {11},
	number = {4},
	pages = {259--263},
	issn = {1749-4893},
	doi = {10.1038/nphoton.2017.33},
	url = {https://doi.org/10.1038/nphoton.2017.33},
	publisher = {Nature Publishing Group},
	language = {en}
}

@article{Wang,
	author = {Zegui Wang and Qijun You and Yun Gao and Peixiang Lu and Wei Cao},
	journal = {Photon. Res.},
	number = {6},
	pages = {1666--1673},
	publisher = {Optica Publishing Group},
	title = {High-precision spatiotemporal profiler of femtosecond laser pulses},
	volume = {13},
	month = {Jun},
	year = {2025},
	url = {https://opg.optica.org/prj/abstract.cfm?URI=prj-13-6-1666},
	doi = {10.1364/PRJ.559295},
}

@article{YOU2,
	title = {Broadband multi-distance lensless imaging},
	journal = {Optics \& Laser Technology},
	volume = {184},
	pages = {112374},
	year = {2025},
	issn = {0030-3992},
	doi = {10.1016/j.optlastec.2024.112374},
	url = {https://www.sciencedirect.com/science/article/pii/S0030399224008323},
	author = {Qijun You and Yun Gao and Fucai Zhang and Qing Liao and Wei Cao and Peixiang Lu},
}
    \bibliographyfullrefs{bibliography}
	
\end{document}